\documentclass[aps,pre,twocolumn,longbibliography,showpacs,10pt]{revtex4-2}

\usepackage[latin1]{inputenc}
\usepackage{amsmath}
\usepackage{amssymb}
\usepackage{bbold}
\usepackage{graphicx}
\usepackage{enumerate}
\usepackage{libertine}
\usepackage{color}
\usepackage{hyperref}

\newcommand{\av}[1]{\langle{#1}\rangle}
\newcommand{\var}[1]{\mathrm{Var}({#1})}
\newcommand{\vmr}{\xi}
\newcommand{\vv}{\mathbf{w}}
\newcommand{\uu}{\mathbf{u}}

\newcommand{\rme}{\mathrm{e}}

\newcommand{\rmd}{\mathrm{d}}

\begin{document}

\title{Clusters determine local fluctuations of random walks on graphs}

\author{M.~Bruderer}
\affiliation{Wenger Engineering GmbH, Research and Development, Einsteinstraße~55, 89077 Ulm, Germany}
\email{martin.bruderer@wenger-engineering.com}

\date{\today}

\begin{abstract}\fontsize{10}{12}\selectfont
The evolution of many stochastic systems is accurately described by random walks on graphs.
We here explore the close connection between local steady-state fluctuations of random walks
and the global structure of the underlying graph. Fluctuations are quantified by the number
of traversals of the random walk across edges during a fixed time window, more precisely,
by the corresponding counting statistics.
The variance-to-mean ratio of the counting statistics is typically lowered if two end vertices of
an edge belong to different clusters as defined by spectral clustering. In particular, we relate
the fluctuations to the algebraic connectivity and the Fiedler vector of the graph. Building on
these results we suggest a centrality score based on fluctuations of random walks. Our findings
imply that local fluctuations of continuous-time Markov processes on discrete state space depend
strongly on the global topology of the underlying graph in addition to the specific transition rates.
\end{abstract}

\maketitle

\fontsize{11}{13.1}\selectfont


\section{Introduction}

The representation of complex systems by graphs has been proven to be extremely useful in areas
spanning physics, biology and economical sciences~\cite{Newman-SIAM-2003,Newman-2010,Estrada-2011}.
Graphs are not only suited to encode structural properties, the time evolution of many systems can
be accurately modeled in terms of random walks on graphs~\cite{Burioni-JPA-2005,Lyons-2005}, for
example, to investigate the performance of communication networks~\cite{Akyildiz2000} or the motion
of molecular motors on microtubules~\cite{Kolomeisky-RPC-2007}. In this setting, the graph describes
the discrete phase space of the system, where vertices and edges represent the states and allowed
transitions, respectively. The time evolution is then equivalent to a continuous-time random walk
on the graph, consisting of successive jumps between states, separated by random dwell times.

An important aspect of such models is the relation between the random walk and the underlying graph.
The inevitable question is how exactly the evolution of the random walk is related to
the graph topology. This question was first addressed in the work of Kirchhoff~\cite{Kirchhoff-APC-1847,Kirchhoff-IRE-1958}
and much later by Hill~\cite{Hill-JTB-1966,Hill-2012} and Schnakenberg~\cite{Schnakenberg-RMP-1976} who
related the steady-state properties of continuous-time random walks to graph elements such as spanning
trees and graph cycles. The initial results have since then been considerably extended in several directions.
Schnakenberg's network theory has been applied, for instance, to classify nonequilibrium steady
states~\cite{Zia-JSM-2007} and to establish fluctuation theorems for nonequilibrium
systems~\cite{Andrieux-JSP-2007}.

\begin{figure}[t]
\centering
\vspace{5pt}
\includegraphics[width=\columnwidth]{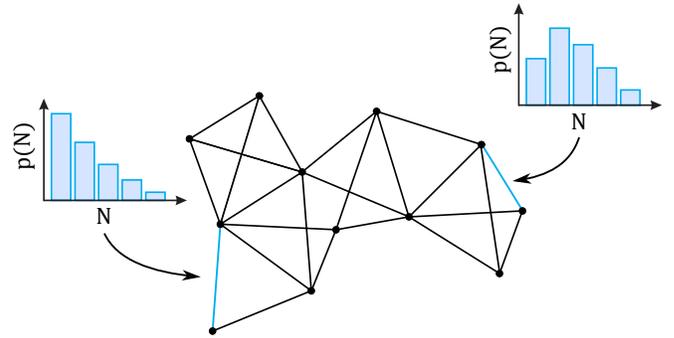}
\caption{The evolution of a system is modeled as a continuous-time random walk on a graph: vertices and edges
represent states and allowed transitions of the system. Local steady-state fluctuations of the random walk are
quantified by the counting statistics $p(N)$ of the number of jumps $N$ across a specific edge during a fixed
time window. The clusters of the underlying graph determine whether the variance-to-mean ratio of the counting
statistics for an edge is either sub-Poissonian or super-Poissonian.}
\label{scheme}
\end{figure}

We here investigate the close connection between the local steady-state fluctuations of random walks and the
global structure of the underlying graph. Our practical approach to the problem mainly builds on spectral clustering
theory~\cite{Luxburg2007} and the established tools of counting statistics~\cite{Flindt-PRB-2010,Bruderer-NJP-2014},
which are used for characterizing the properties of random walks. We focus on the partition of the graph into clusters,
i.e.~subgraphs with highly connected vertices~\cite{Fortunato2010}, where clusters are precisely defined by
spectral clustering.

Random walks on graphs have already been exploited to define betweenness centralities~\cite{Newman-SN-2005},
to characterize the similarity between vertices of graphs~\cite{Fouss2007} and to detect
clusters~\cite{harel2001clustering,zhang2016limited,lucinska2016graph,okuda2019community}.
Specifically, centrality scores~\cite{Borgatti-SN-2005,Koschutzki-2005} with the goal of identifying the most important
vertices or edges of graphs are closely related to the detection of clusters~\cite{Girvan-PNAS-2002,Newman-PRE-2004}.
In contrast to our approach, however, most of these results are not based on steady-state properties of random walks.

The local fluctuations of the random walk are specified for each edge by the number of jumps $N$ across
the edge during a fixed time window and summarized by the counting statistics for edge traversals $p(N)$,
as illustrated in Fig.~\ref{scheme}. We mainly consider the idealized case where all transition rates
between states are equal to unambiguously identify the influence of the graph structure and to emphasize
the role of fluctuations on top of average quantities, which in contrast are independent of the graph
structure.

Our analysis allows us to establish direct relations between local fluctuations of random walks and clusters of
the graph. As an example, for graphs with two dominating clusters we may write in a broad sense
\begin{equation}\nonumber	
	\rm{edge}\;\rm{fluctuations}\sim\frac{\rm{cluster}\;\rm{membership}}{\rm{algebraic}\;\rm{connectivity}}\,,
\end{equation}
where the cluster membership determines the intensity of the fluctuations. If the end vertices of an edge are
members of different clusters then the fluctuations are generally reduced to sub-Poissonian values.

Building on our results we put forward a new centrality score, referred to as fluctuation centrality,
for identifying the most important edges of a graph, a fundamental and still unsettled problem
in network analysis. The score assigns to each edge $\ell=(i,j)$ between vertices $i$ and $j$ the value
$\sigma_{FC}(\ell)  = 1/(1 + 2L^{+}_{i,j})$, where $L^{+}$ is the Moore-Penrose inverse of the Laplacian
matrix $L$. The distinguishing feature of the fluctuation centrality is that, in principle, it can be
measured in physical systems that are described by random walks on graphs.

In the first part of this paper we establish the exact relation between the fluctuations of random walks
and the structure of the graph and present a few basic applications. In the second part we define the
fluctuation centrality score and compare it to the established betweenness centrality. Finally, we discuss
possible extensions of our approach and implications of our results in the conclusions.


\section{Graph structure and fluctuations}

We consider an unweighted undirected graph $G=(E,V)$ that consists of a finite set of vertices $V$ connected
by a set of edges $E$, where $(i,j)\in E$ denotes an edge between the vertices $i, j\in V$ and $n=\vert V\vert$
is the number of vertices. The graph is assumed to be connected and has neither multiple edges nor self-loops.
Commonly used matrix representations of the graph are the adjacency matrix $A$ with entries
\begin{equation}
	 A_{ij} =
  \begin{cases}
   1 & \text{if } (i,j)\in E \\
   0 & \text{otherwise}
  \end{cases}
\end{equation}
and the Laplacian (or Kirchhoff) matrix $L = D - A$. The degree matrix $D$ is diagonal with
$D_{ii}=d_i$, where $d_i$ is the number of edges connected to vertex~$i$.

The eigenvalues of $L$ form the graph spectrum, which is essential for spectral clustering.
The eigenvalues are labelled such that $\lambda^{(1)}\leq\lambda^{(2)}\leq\cdots\leq\lambda^{(n)}$ and
$\uu^{(1)},\uu^{(2)},\ldots,\uu^{(n)}$ are the corresponding right eigenvectors. The smallest eigenvalue $\lambda^{(1)} = 0$
is the unique zero eigenvalue with the corresponding eigenvector $\uu^{(1)} = (1,\ldots,1)^{\rm T}/\sqrt{n}$.
Particularly important are the second smallest eigenvalue $\lambda^{(2)}$, known as the algebraic connectivity
of $G$~\cite{de2007old}, and the corresponding eigenvector $\uu^{(2)}$, called Fiedler vector~\cite{Fiedler1973,Fiedler1975}.
We assume that all right eigenvectors are normalized such that $[\uu^{(k)}]^{\rm T}\uu^{(k)}=1$.

The central idea of spectral clustering is that two vertices $i$ and $j$ belong to the same cluster if the components
$u^{(k)}_i$ and $u^{(k)}_j$ of the eigenvector $\uu^{(k)}$ have comparable values~\cite{Luxburg2007}. In practice,
vertices are partitioned into clusters based on a small set of eigenvectors $\uu^{(k)}$. Already the algebraic connectivity
and the Fiedler vector provide substantial information about the structure of the graph. More precisely,
partitioning graphs into two clusters based on the components of the Fiedler vector $\uu^{(2)}$ is known as the
spectral bisection method, i.e.~two clusters are defined by vertices for which either $u^{(2)}_i \geq 0$ or $u^{(2)}_i < 0$.

\subsection{Fluctuations of random walks}

Apart from being a matrix representation, the Laplacian $L$ is also the generator of the time evolution
of the continuous-time random walk on $G$, restricted to the Markovian case. This type of random walk is
commonly called edge-centric as opposed to vertex-centric~\cite{masuda2017random}. The values of the
off-diagonal entries of $L$ are rates, all identical, for the transitions between different vertices. On
the level of average quantities, the evolution is described by the occupation probabilities $w_i(t)$ that
the random walk is found on vertex $i$ at time $t$. The  probabilities $w_i(t)$ are determined by the
master equation
\begin{equation}\label{randomwalk}
	\frac{\rmd \vv(t)}{\rmd t} = -L\vv(t)\,,
\end{equation}
with the column vector $\vv = (w_i,\ldots,w_n)^{\rm T}$. For the case of connected graphs, the eigenvector $\uu^{(1)}$ is the
unique steady-state solution defined by the equation $L\vv^{\rm s} = 0$. The random walk in the steady state is
therefore found on each vertex $i$ with equal probability $w^{\rm s}_i=1/n$ and the occupations $w^{\rm s}_i$ are
therefore independent of the graph structure.

Local fluctuations of the steady state are manifest on the level of individual trajectories, consisting of
successive jumps along graph edges separated by random dwell times on vertices. To quantify local fluctuations
we monitor the jumps between neighboring vertices: For an edge $\ell=(i,j)\in E$ connecting the vertices
$i, j\in V$ we count the number of times $N_\ell$ that the random walk jumps from $i$ to $j$ during a fixed
time window $\tau$ of sufficient length. The number of traversals $N_\ell$ is a random variable associated with
each edge $\ell$ and described by the probability distribution $p_\ell(N)$, the counting statistics for
transitions along graph edges.

We identify the steady-state fluctuations of the random walk at edge $\ell$ with the variance-to-mean ratio
$\vmr_\ell\equiv\var{N_\ell}/\av{N_\ell}$. For a Poissonian distribution, taken as a reference here, mean
and variance are equal and therefore $\vmr_\ell = 1$. Deviations from the Poissonian value are referred to
as sub-Poissonian $\vmr_\ell < 1$ and super-Poissonian $\vmr_\ell > 1$. The mean $\av{N_\ell}$, similar to
the occupations $w_i$, is identical for all edges. We will show that local fluctuations, quantified by
$\vmr_\ell$, depend sensitively on the structure of the graph, as illustrated in Fig.~\ref{cluster}.

Our goal is to express the variance-to-mean ratio in terms of the eigenvalues $\lambda^{(k)}$ and eigenvectors
$\uu^{(k)}$, which in turn are related to the clusters of $G$. To this end we follow the established theory of
counting statistics~\cite{Flindt-PRB-2010,Bruderer-NJP-2014} and apply known results on parameter-dependent
eigenvalues~\cite{Lancaster1964}. We define for each edge $\ell$ the modified adjacency matrix $A_\ell(\theta)$
with entries
\begin{equation}
	 [A_\ell]_{ij}(\theta) =
  \begin{cases}
   \rme^\theta A_{ij} & \text{if } (i,j)=\ell\\
   A_{ij} & \text{otherwise}
  \end{cases}
\end{equation}
and the Laplacian $L_\ell(\theta) = D - A_\ell(\theta)$, both depending on the edge $\ell$ and the parameter $\theta$.
The parameter-dependent matrices $A_\ell(\theta)$ and $L_\ell(\theta)$ are identical to their unmodified counterparts
$A$ and $L$ in the limit $\theta\rightarrow 0$. We denote by $\lambda_\ell(\theta)$ the parameter-dependent eigenvalue
corresponding to the zero eigenvalue $\lambda^{(1)}$ of the Laplacian, that is $\lambda_\ell(\theta) =\lambda^{(1)}$
in the limit $\theta\rightarrow 0$. The eigenvalue $\lambda_\ell(\theta)$ is related to the cumulant generating function
$g_\ell(\theta) = \log\av{\rme^{\theta N_\ell}}$ through the identity $g_\ell(\theta)/\tau = \lambda_\ell(\theta)$.
All cumulants $C_{k,\ell}$ of order $k$ of the probability distribution $p_\ell(N)$ are found from
\begin{equation}\label{eigencm}
	\frac{C_{k,\ell}}{\tau} = \frac{\partial^k \lambda_\ell(\theta)}{\partial \theta^k}\bigg|_{\theta=0}\!.
\end{equation}
The cumulants therefore quantify how sensitive the graph invariant $\lambda^{(1)}$ is to infinitesimally small changes
in the weight of the edge $\ell$.

The first two cumulants are simply the mean and variance of $N_\ell$, i.e., $C_{1,\ell} = \av{N_\ell}$ and
$C_{2,\ell} = \var{N_\ell}$. The explicit evaluation of the derivatives of the parameter-dependent eigenvalue
$\lambda_\ell(\theta)$ yields, as a main result of the paper, the mean value $\av{N_\ell} = \tau/n$ and
variance-to-mean ratio
\begin{equation}\label{vmr_result}
	\vmr_{i,j} = 1 + 2\sum_{k=2}^n\frac{u^{(k)}_i u^{(k)}_j}{\lambda^{(k)}}\qquad\ell=(i,j)\,.
\end{equation}
The variance-to-mean ratio $\vmr_{i,j}$ at edge $\ell=(i,j)$ is either sub- or super-Poissonian, depending on
the sum on the right hand side of equation~\eqref{vmr_result}. If the end vertices $i$ and $j$ belong to the
same cluster then the product $u^{(k)}_i u^{(k)}_j$ is positive. However, if vertices $i$ and $j$ belong
to different clusters then the product can take negative values.

The situation is particularly transparent if the algebraic connectivity $\lambda^{(2)}$ is much smaller than
all other eigenvalues such that
\begin{equation}\label{vmr_fiedler}
	\vmr_{i,j} \approx 1 + 2\frac{u^{(2)}_i u^{(2)}_j}{\lambda^{(2)}}\qquad\ell=(i,j)\,.
\end{equation}
The correction to the Poissonian variance-to-mean ratio can be substantial because the algebraic connectivity
$\lambda^{(2)}$ may take small values even for large graphs~\cite{de2007old}. This is made explicit by the
general inequality $2\lambda(G)[1-\cos(\pi/n)]\leq\lambda^{(2)}\leq\kappa(G)$, where $\lambda(G)$ and $\kappa(G)$
are the edge and vertex connectivities, respectively~\cite{Cvetkovic-1980}. The product $u^{(2)}_i u^{(2)}_j$ of
the components of the Fiedler vector, which partitions the graph into two dominating clusters, is negative
for vertices that belong to different clusters and positive for vertices that are member of the same cluster.

It is convenient, e.g.~for numerical implementations, to write equation~\eqref{vmr_result} in terms of the Moore-Penrose
inverse $L^{+}$ of the Laplacian matrix $L$~\cite{Fouss2007}, namely
\begin{equation}
	\vmr_{i,j} = 1 + 2L^{+}_{i,j}.
\end{equation}
Both the Laplacian $L$ and thus the matrix $L^{+}$ are symmetric and it follows immediately that for unweighted
undirected graphs $\vmr_{i,j} = \vmr_{j,i}$.

Other interesting graph quantities such as the average first-passage time of randoms walks~\cite{Fouss2007} and the
resistance distance~\cite{klein1993resistance} can also be expressed in terms of the Moore-Penrose inverse. More
importantly, the topological centrality~\cite{ranjan2013geometry} is closely related to the fluctuation centrality
as discussed in following section. Calculating the Moore-Penrose inverse $L^{+}$ for a graph with $n$ vertices is
equivalent to finding the singular value decomposition~(SVD) of the Laplacian $L$ with a cost of $O(n^3)$
floating-point operations~\cite{Golub-SIAM-1965}.

\subsection{Fluctuations of Markov processes}

The previous results can be further adapted to time-continuous Markov processes. For such processes the
transitions are not necessarily bidirectional and characterized by specific jump rates. This can be naturally
accounted for by extending the previous description to directed graphs with weighted edges. The random walk
on the graph and time-continuous Markov processes are then equivalent~\cite{Mirzaev-BMB-2013}.

The weighted adjacency matrix $A$ has entries $A_{ij} = r_{ij}$ if $(i,j)\in E$ and $A_{ij} = 0$ otherwise,
where $r_{ij}$ denote the jump rates of the Markov process. The degree matrix $D$ is diagonal with
$D_{ii}=\sum_{j\neq i} r_{ij}$ and Laplacian matrix is $L = D - A$. Since the Laplacian $L$ is not
necessarily symmetric the right eigenvectors $\uu^{(k)}$ may differ from the left eigenvectors, which
are denoted by $\tilde{\uu}^{(k)}$ and normalized such that $\tilde{\uu}^{(k)}\uu^{(k)}=1$.

For the case of time-continuous Markov processes, restricted to bidirectional transitions,
we then obtain the mean value
\begin{equation}\label{mean_general}
	\av{N_\ell} = \tau\,\tilde{u}^{(1)}_i r_{ij} u^{(1)}_j\qquad\ell=(i,j)\,
\end{equation}
and variance-to-mean ratio
\begin{equation}\label{vmr_general}
	\vmr_{i,j} = 1 + 2\sum_{k=2}^n\frac{\tilde{u}^{(k)}_i r_{ij} u^{(k)}_j}{\lambda^{(k)}}\qquad\ell=(i,j)\,.
\end{equation}
The specific jump rates $r_{ij}$ modify both the mean value $\av{N_\ell}$ and the fluctuations $\vmr_{i,j}$,
which for random walks with equal rates are determined entirely by the graph structure. The fluctuations
observed in time-continuous Markov processes are thus partly determined by the specific transition rates and
partly by the graph structure of the discrete phase space.

\subsection{Applications}

We present for concreteness a few graphs and the corresponding fluctuations of the random walk. An instructive
example, where equation~\eqref{vmr_fiedler} is exact, is the graph consisting of two vertices connected by a single
edge. The algebraic connectivity is $\lambda^{(2)} = 2$, the Fiedler vector is  $\uu^{(2)} = (-1,1)^{\rm T}/\sqrt{2}$
and the parameter-dependent eigenvalue $\lambda_\ell(\theta) = 1 + \exp(\theta/2)$. Either from equation~\eqref{eigencm}
or equation~\eqref{vmr_fiedler} one finds the sub-Poissonian fluctuations $\vmr_{1,2} = 1/2$.

\begin{figure}[t]
\centering
\includegraphics[width=200pt]{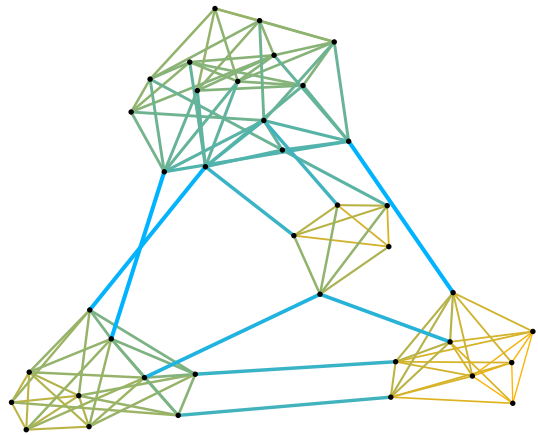}
\caption{A graph composed of four densely connected clusters: The steady-state fluctuations of the random walk along the
bridge-like edges are sub-Poissonian wheres fluctuations within clusters are super-Poissonian, indicated by blue (thick)
lines and yellow (thin) lines, respectively. The variance-to-mean ratio of the counting statistics for edge traversals
ranges from $0.96$ to $1.46$ and the smallest four eigenvalues of the Laplacian are $0, 0.37, 0.59, 0.90$.}
\label{cluster}
\end{figure}

The previous result generalizes qualitatively if the two vertices are each replaced by a cluster
(cf.~Fig.~\ref{channel}) and even if a graph is composed of several dense clusters. Let us consider
several clusters that are connected by a few bridge-like edges, as exemplified by the random graph
in Fig.~\ref{cluster}. We see indeed that the fluctuations of the random walk along the bridge-like
edges are reduced to sub-Poissonian values. Note that for a graph with $m$ clusters that are only weakly
connected the eigenvalues $\lambda^{(2)},\ldots,\lambda^{(m)}$ are small. As a consequence, terms
weighted by the factors $1/\lambda^{(2)},\ldots,1/\lambda^{(m)}$ contribute substantially to the sum in
equation~\eqref{vmr_result}. Thus, only a limited set of eigenvectors is relevant for the fluctuations
in strongly clustered graphs.

Random walks on highly symmetric graphs, e.g.~the complete graph, star graph or cycle graph, exhibit identical fluctuations
on all edges. We therefore focus on the scaling of the fluctuations with the number of vertices $n$. The fluctuations for
a random walk on the complete graph and star graph are identical and given by $\vmr_{i,j} = 1 - 2/n^2$~\cite{ranjan2014incremental},
i.e.~in the limit of large graphs the fluctuations approach a Poissonian variance-to-mean ratio. In contrast, the fluctuations
for the cycle graph scale as $\vmr_{i,j} = n/6 + 5/(6n)$ and therefore grow linearly with $n$ for large graphs. Generally,
these symmetric graphs have highly degenerate spectra so that none of the contributions in equation~\eqref{vmr_result} can
be neglected. For instance, the non-zero eigenvalues of the complete graph are all identical, namely $\lambda^{(k)} = n$
for $k\geq2$.


\section{Fluctuation centrality score}

The previous findings explicitly show that the graph structure strongly influences local fluctuations
of random walks. We can therefore learn about the structural role of an edge from the observed fluctuations.
A way to implement this idea is to introduce a fluctuation-based centrality score that identifies the most
important edges of the graph. Such a score $\sigma$ is defined as a mapping from the set of edges~$E$ (or
vertices $V$) to the real numbers $\mathbb{R}$, where high scores indicate a high centrality of the edge
(or vertex)~\cite{Koschutzki-2005}. Commonly used centrality measures are, for instance, the betweenness
centrality~\cite{Girvan-PNAS-2002,Newman-PRE-2004}, the eigenvector centrality~\cite{Bonacich-JMS-1972}
and closeness centrality~\cite{Freeman-SN-1979}.

The fluctuation centrality defined here assigns to each edge $\ell$ the centrality score
${\sigma_{FC}}(\ell) = 1/\xi_\ell$ or equivalently
\begin{equation}
	\sigma_{FC}(\ell)  = \frac{1}{1 + 2L^{+}_{i,j}}\qquad\ell=(i,j).
\end{equation}
The fluctuation centrality is thus defined as the mean-to-variance ratio of the counting statistics for
transitions along graph edges and takes positive values for all edges. Edges with a high fluctuation
centrality connect different clusters, whereas a low fluctuation centrality is typical for edges within
clusters. Note that alternatively any monotonically decreasing function of $\xi_\ell$ may be used for
the definition of a fluctuation-based centrality score.

The fluctuation centrality for edges, although motivated differently, is very similar
to the topological centrality for nodes, defined for node $i$ by~\cite{ranjan2013geometry}
\begin{equation}
	\sigma_{TC}(i) = \frac{1}{L^{+}_{i,i}}\,.
\end{equation}
Unlike the off-diagonal elements, the diagonal elements of the Moore-Penrose inverse
$L^{+}_{i,i} = \sum_{k=2}^n[u^{(k)}_i]^2/\lambda^{(k)}$ take always positive values.
In the same way as the fluctuation centrality, the topological centrality of a node
is a function of the entire spectrum of the Laplacian $L$.

Let us finally compare the fluctuation centrality to the well-known edge betweenness centrality~\cite{Girvan-PNAS-2002,Newman-PRE-2004}.
This quantity also captures the global graph structure and measures the number of shortest paths between
all vertices that pass through the edge of interest. If we denote by $\gamma_{i,j}$ the number of shortest
paths between vertices $i$ and $j$ and by $\gamma_{i,j}(\ell)$ the number of shortest paths that contain
the edge $\ell$ then the edge betweenness is defined as
\begin{equation}
	{\sigma_{EB}}(\ell) = \sum_{i,j\in V}\frac{\gamma_{i,j}(\ell)}{\gamma_{i,j}}\,.
\end{equation}
An edge with a high edge betweenness $\sigma_{EB}(\ell)$ represents a central connection between
different parts of a graph.

\begin{figure}[t]
\centering
\raisebox{2.3cm}{(a)}\hspace{10pt}\includegraphics[width=195pt]{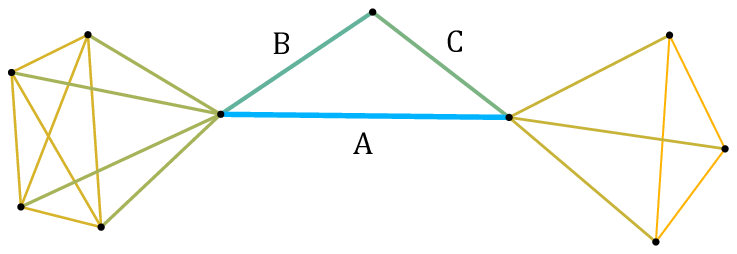}\vspace{10pt}\\
\raisebox{2.3cm}{(b)}\hspace{10pt}\includegraphics[width=195pt]{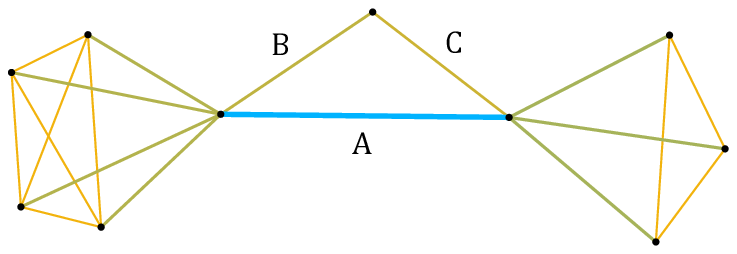}\vspace{0pt}
\caption{The fluctuation centrality~(a) in comparison to the edge betweenness~(b) for two clusters
connected by bridge-like edges. Blue (thick) edges correspond to high centrality and yellow (thin)
edges to low centrality scores. The fluctuation centrality gives high scores to all central edges ${\tt A}$,
${\tt B}$ and ${\tt C}$, while the edge betweenness only singles out edge ${\tt A}$. The scaled
scores for edges ${\tt A}$, ${\tt B}$, ${\tt C}$ are $1.0, 0.61, 0.51$ and $1.0, 0.21, 0.16$ for
(a) and (b), respectively.}
\label{channel}
\end{figure}

Unlike the edge betweenness, the fluctuations centrality assigns high scores to edges that are not part
of a shortest path, but still act as important connectors between clusters. This is illustrated by the
example of two clusters linked by a few central edges~\cite{Newman-SN-2005}, shown in Fig.~\ref{channel}.
The edge betweenness identifies edge ${\tt A}$ as most important while giving low scores to edges ${\tt B}$
and ${\tt C}$ (and all other edges). In contrast, the fluctuation centrality assigns high scores to edges
${\tt A}$, ${\tt B}$ and ${\tt C}$, which agrees with our intuition that all three edges are important
connections between the two clusters.


The similarity to the betweenness centrality allows us to use the original Girvan-Newman
algorithm~\cite{Girvan-PNAS-2002,Newman-PRE-2004} in combination with the fluctuation
centrality to detect clusters. The algorithm detects clusters by progressively removing
edges such that the remaining components of the graph are highly connected. In its basic
form the algorithm corresponds to finding the highest fluctuation centralities in order
to identify the two dominating clusters of the graph.

A distinguishing feature of the fluctuation centrality is the fact that, in principle, it can be measured in certain
physical systems, which are described by random walks on graphs. Measuring the variance-to-mean ratio of only a few
edges or even a single edge may already provide valuable information about the system. In particular, sub-Poissonian
fluctuations at a single edge indicate that the edge is placed between clustered parts of the underlying graph. 


\section{Conclusions}

We have established a quantitative relation between steady-state fluctuations of random walks and the clusters
of the underlying graph. An essential insight provided by our results is that \emph{local} fluctuations are
determined by the \emph{global} structure of the graph, which is quantified by the fluctuation centrality score. 
The random walk in the steady state explores the entire graph and retrieves structural information encoded in
local statistical properties.

Stochastic numerical simulation, i.e.~sampling over trajectories, is an alternative and practical way to determine
the fluctuations of the random walk and the fluctuation centrality score of edges. GraphStream, for example, is a Java
library that allows the user to keep track of the number of edge traversals~\cite{Pigne2008}. The run time of the
stochastic simulation has to be significantly longer than the cover time, i.e.~the expected time taken for a random
walk to visit every vertex at least once~\cite{Kahn1989}.

Our results are relevant to Markov processes, equivalent to random walks on graphs with weighted edges. While the
generalization to weighted graphs may obscure the precise role of the graph structure it covers a large class of
physical systems described by classical master equations. The fluctuations in such systems thus are always
determined, in principle, by both the transition rates and the structure of the discrete phase space.

The procedure presented in this paper is applicable to generators of alternative versions of random walks on graphs,
either discrete or continuous in time. For any generator, the counting statistics for edge traversals is found from
a parameter-dependent eigenvalue and therefore related to the eigenvectors of the generator, which in turn contain
information about the structure of the graph. In particular, continuous-time \emph{quantum} stochastic walks on
graphs may reveal information that is not accessible to classical random walks~\cite{Bruderer2016}.


\bibliography{cluster_paper_arxiv_2}

\end{document}